\documentstyle[12pt]{article}
\def\be{\begin{equation}}
\def\bd{\begin{displaymath}}
\def\ee{\end{equation}}
\def\ed{\end{displaymath}}
\begin{document}
\thispagestyle{empty}


\vskip 1.5cm
\begin{center}
{\Large\bf Coupling Constant Unification and LEP Data}
\vskip 2cm
Hang Bae Kim and Jihn E. Kim

\sl Center for Theoretical Physics and Department of Physics\\
Seoul National University\\
Seoul 151-742, Korea
\rm
\end{center}
\vskip 3.5cm
{\center Abstract\\}

The recent LEP data for gauge coupling constants constrain many grand
unified models.  In this paper,
we study several possibilities for unification of gauge coupling
constants.  Without an intermediate scale, the minimal
supersymmetric standard model with two Higgs doublets is the only
possibility.  For one intermediate scale,
we present a few unification schemes without supersymmetry.

\newpage

One of the best theoretical ideas in the last few decades has been to
understand the strengths of gauge coupling constants.  The first great
advance in this direction has been the invention of grand
unified theories\cite{gut}.  The second advance has been four
dimensional superstring models\cite{4ds}.\footnote{The ten dimensional
heterotic string\cite{het} unifies coupling constants, but one can say
that it is unified
because of the grand unification group $E_8\times E_8^\prime$.}
The string theory requires that the coupling must be the same at the
string scale\cite{weinberg}.

Among the predictions of GUT, the value $\sin^2\theta_W$, proton decay and
fermion mass ratios have attracted most attention.
Proton decay experiment constrained the unification scale\cite{proton}.
In this regard, LEP data\cite{coupling} of
three coupling constants have played the crucial role in testing this idea.
Until recently, the strong coupling has not been measured accurately,
and the value of $\sin^2\theta_w$ was the prime constraint for the
unification models.  Thus the accurate measurement of $\alpha_c$ at
LEP\cite{bet} triggered an interest for the study of unification
condition.  For example,
Giveon, Hall and Sarid \cite{ghs91}
studied the criteria for unification of
coupling constants recently.  Our philosophy here is the same as
their's: {\it in search of greater number of possible unification
models.}  With a more accurate determination of $\alpha_c$
available now\cite{bet}, it
is timely to study this problem again. Furthermore, in this paper
we go beyond the minimal unification models by introducing
an intermediate mass scale.

The success of coupling constant unification originates from the
observation that the apparent difference of coupling constants at low
energy is attributed to the running of coupling constants\cite{gqw}.
If the coupling constants are unified to $\alpha_X$ at some high energy
scale, say $M_X$, then it evolves to
\begin{equation}
\alpha_i^{-1}(\mu)\ =\ \alpha^{-1}_X-{b_i\over 2\pi}\ln\left(
{\mu\over M_X}\right)
\end{equation}
where $\mu$ is the scale in question, and $b_i$ is the standard notation
for the coefficient of $\beta_i$.  Coupling constants in Eq.~(1) are
defined for normalized generators.  For the electroweak hypercharge, we
use $\alpha_Y$ for the usual coupling and $\alpha_y$ for the normalized
hypercharge; thus $\alpha_y={5\over 3}\alpha_Y$.
Then the difference of coupling constants below $M_X$ but above
a new physics scale $M_I$ satisfies
\begin{equation}
\alpha_i^{-1}-\alpha_j^{-1}\ =\ -{b_i-b_j\over2\pi}\ln\left(\mu\over M_X\right)
\end{equation}
whence we obtain
\begin{equation}
\label{criterion}
{\alpha_i^{-1}(\mu)-\alpha_j^{-1}(\mu)\over
 \alpha_j^{-1}(\mu)-\alpha_k^{-1}(\mu)}\ \equiv\ {b_i-b_j\over b_j-b_k}.
\end{equation}
Eq.~(\ref{criterion}),
which must hold independent of scale $\mu$, is a one-loop
{\it criterion for successful unification of coupling constants} at some scale.
Suppose there are {\it two interesting mass scales},
the unification scale $M_X$ and the electroweak scale $M_Z$.
Then the left-hand side can be evaluated by data at the scale $M_Z$.
On the other hand, the right-hand side is calculated
in a specific model for unification.
If they turn out to be the same within experimental and theoretical errors,
the model is not in conflict with low energy data.
If they differ, the model is ruled out.
The use of Eq.~(\ref{criterion}) is simplified since the right-hand side
usually does not depend on the fermion content of the theory.
It is mainly determined by the gauge group and the Higgs content corrects it
by small amount.  The specific role of the Higgs fields is due to the
assumption on the split multiplet of Higgs fields for proton stability.

If no new physics scale is present between $M_X$ and $M_Z$,
the unification mass is given by
\begin{equation}
M_X\ =\ M_Z\exp\left[2\pi{\alpha_w^{-1}(M_Z)-\alpha_c^{-1}(M_Z)\over
b_w-b_c}\right]
\end{equation}
where $\alpha_w$ and $\alpha_c$ are $SU(2)_L$ and $SU(3)_c$ couplings, and
$b_w$ and $b_c$ are coefficients of the respective $\beta$'s.
For a successful unification, proton lifetime requires $M_X>10^{15\sim
16}$ GeV, which gives another constraint
\begin{equation}
b_w-b_c<(2\pi \log_{10}e){\alpha_w^{-1}(M_Z)-\alpha_c^{-1}(M_Z)\over
(15\sim 16)-\log_{10}(M_Z/{\rm GeV})} \label{plc}
\end{equation}

As the first example, let us consider the possibility of unification of
the standard model. Let us define
\begin{equation}
r(\mu)=\frac{\alpha_y^{-1}-\alpha_w^{-1}}{\alpha_w^{-1}-\alpha_c^{-1}}
,\hspace{10mm}
R=\frac{b_y-b_w}{b_w-b_c}
\end{equation}
{}From the LEP data\cite{coupling}
\begin{equation}
\begin{array}{rcl}
\alpha_y^{-1}(M_Z)&=&58.9\pm 0.3\ \\
\alpha_w^{-1}(M_Z)&=&29.7\pm 0.2\ \\
\alpha_c^{-1}(M_Z)&=&8.47\pm 0.5\
\end{array}
\end{equation}
we obtain
\begin{equation}
r(M_Z) = 1.37 \pm 0.07
\end{equation}
Let $h_2$ be the number of Higgs doublets and $n_g$ be the number of
generations. Then, we have
\begin{equation}
\begin{array}{rcl}
b_y &=& {4\over 3}n_g+{1\over 10}h_2,\\
b_w &=& -{22\over 3}+{4\over 3}n_g+{1\over 6}h_2,\\
b_c &=& -11+{4\over 3}n_g
\end{array}
\end{equation}
Thus,
\begin{equation}
R\ =\ {{22\over 3}-{1\over 15}h_2\over {11\over 3}+{1\over 6}h_2}\ =\
2\ (h_2=0)\ ,\ \ 1.90\ (h_2=1)\ ,\ \ 1.36\ (h_2=8)
\end{equation}
Thus the coupling constant unification does not occur in the minimal
($h_2=1$) standard model.  See Fig.~1 and 2(a).  In Fig.~1, the
horizontal lines
correspond to $R$ and three curly lines correspond to experimentally
determined $r(\mu)$'s within $1\sigma$.   The crossing point should be
$M_Z$.   Introduction of enough Higgs doublets ($h_2=8$) at low
energy makes the theory unifiable, because they can meet at
$\mu\sim M_Z$. But the proton lifetime constraint is not satisfied;
the RHS of Eq.~(\ref{plc}) is 4.44 $\sim$ 4.13 and
the LHS of Eq.~(\ref{plc}) is $11/3+h_2/6=5$.  In Fig.~2, we show
the evolution of coupling constants in the standard model and in
the minimal supersymmetric standard model.

As the second example, let us consider the supersymmetric standard model.
For simplicity we assume that the supersymmetry breaking scale is comparable
to the electroweak scale.  Then, we can use
\begin{equation}
\begin{array}{rcl}
b_y &=& 2n_g+{3\over 10}h_2,\\
b_w &=& -6+2n_g+{1\over 2}h_2,\\
b_c &=& -9+2n_g,
\end{array}
\end{equation}
and obtain
\begin{equation}
R\ =\ 2\ \ (h_2=0) ,\ \ 1.40 \ \ (h_2=2)
\end{equation}
Thus the coupling constant unification is successful in the minimal
supersymmetric standard model with $h_2=2$.  Because the supersymmetry
breaking scale is very close to the electroweak scale, our study for
the two scale physics for the supersymmetric standard model is
approximately valid.  For a more accurate calculation, we must use
the three scale physics, which is shown in Fig.~2(b) for the
supersymmetry breaking scale $M_{S}=1$ TeV.
Supersymmetric standard models\cite{kim} and $SU(5)\times U(1)$
models\cite{ellis} from superstrings belong to this category.
String theory gives the condition for equal couplings for each gauge group
at the string scale.

Thus the criterion for unification of coupling constants can be
satisfied by extending the minimal model, either by increasing the number
of Higgs doublets or by supersymmetrizing the model.  One may argue that
the LEP data favors the supersymmetric standard model.  However, if
one is forced to introduce many Higgs doublets either from experimental
discovery or from theoretical reasoning of understanding fermion mass
matrix, this argument is no longer valid.   But, in this case the proton
stability must be explained by introducing a symmetry\cite{segre}.
At present, we can conclude
that there are a few paths toward coupling constant unification.

Another logical possibility is the presence of three or
more mass scales; namely we can introduce intermediate mass scales.
The invisible axion idea requires an intermediate scale around $10^{12}$
GeV.  Possibility of lepton number violation needs another intermediate
mass scale.  Grand unifications beyond $SU(5)$ require another scale in
principle.  Therefore, let us introduce intermediate scales for
unification of gauge coupling constants.
For a concrete study, let us introduce just one intermediate
mass scale $M_I$ as the vacuum expectation value of some Higgs fields.
If the vacuum expectation value in question is neutral under the gauge
group at $M_I$, the conclusion is the same as the two scale case studied
above.  Therefore, for the study of three scale cases, let us consider
at the intermediate scale $M_I$
a gauge group $G_I$ {\it which contains the standard model as a proper
subgroup.}

A supersymmetric standard model with two Higgs doublets already satisfies
the unification condition, and we will not consider supersymmetric cases
with intermediate scales.

Without supersymmetry, one may wish to satisfy the unification condition
with a small number of needed Higgs fields by introducing an intermediate
mass scale.  In the remainder of this paper, we will consider this case.
As an example, consider $SO(10)$.  If the symmetry breaking proceeds
via $SO(10)\rightarrow SU(5)_{GG}\times U(1)$ where $SU(5)_{GG}$ is
Georgi and Glashow's $SU(5)$, we redefine the grand unification group
as the $SU(5)_{GG}$.  Then there is no intermediate scale.  On the
other hand, if the symmetry breaking proceeds via $SO(10)\rightarrow
SU(5)_{flipped}\times U(1)$ or $SU(3)\times SU(2)_L\times SU(2)_R
\times U(1)$, then we need an intermediate mass scale to obtain the
standard model at $M_I$.

Let the gauge group be
\begin{equation}
\left\{\begin{array}{cc}
G_3\times G_2\times G_1            & (M_I<\mu<M_X) \\
SU(3)_c\times SU(2)_w\times U(1)_Y & (M_Z<\mu<M_I)
\end{array}\right. ,
\end{equation}
where
\begin{equation}
G_3\supset SU(3)_c,\ \ G_2\supset SU(2)_w
\end{equation}
The electroweak hypercharge generator $Y$ is a combination of a few
generators above $M_I$,
\begin{equation}
Y\ =\ c_1Y_1+c_2Y_2+c_3Y_3
\end{equation}
where $c_i\ (i=1,2,3)$ are numbers and $Y_i$ are normalized
generators belonging to the group $G_i$.
Let $\alpha_i$ be coupling constants of the group $G_i$.
Then, at $M_I$ the hypercharge coupling satisfies at lowest order,
\begin{equation}
{5\over 3}\alpha_y^{-1}\ =\ c_1^2\alpha_1^{-1}+c_2^2\alpha_2^{-1}
+c_3^2\alpha_3^{-1}
\end{equation}
For a unification, the following condition must be satisfied at $\mu=M_I$,
\begin{equation}
r(\mu=M_I)=R
\end{equation}
where
\begin{equation}
r(\mu)={c_1^{-2}[{5\over3}\alpha_y^{-1}-c_2^2\alpha_w^{-1}-c_3^2\alpha_c^{-1}]
-\alpha_w^{-1}\over \alpha_w^{-1}-\alpha_c^{-1}}
\end{equation}
and
\begin{equation}
R = \left.{b_1-b_2\over b_2-b_3}\right|_{\mu=M_I\sim M_X}.
\end{equation}
The common point of $r(\mu)$ and $R$ determines $M_I$.
The proton lifetime constraint can be given as
\begin{equation}
b_2-b_3\ <\ (2\pi\log_{10}e){\alpha_w^{-1}(M_I)-\alpha_c^{-1}(M_I)\over
(15\sim 16)-\log_{10}(M_I/{\rm GeV})}
\end{equation}

As a successful example, let us consider the following symmetry breaking
pattern of $SO(10)$ model,
\begin{eqnarray}
SO(10) & \stackrel{M_X}{\longrightarrow} &
         SU(3)_C\times SU(2)_L\times SU(2)_R \times U(1)_{B-L} \nonumber\\
       & \stackrel{M_I}{\longrightarrow} &
         SU(3)_C\times SU(2)_L\times U(1)_Y
\end{eqnarray}
Since $\alpha_L=\alpha_R\equiv\alpha_2$ for $M_I<\mu<M_X$,
we can apply above formulae with $G_2=SU(2)_L\times SU(2)_R$.
Then,
\begin{equation}
R = \frac{{22}-h_2+{7}h_3}
         {{11}+h_2+{2}h_3}
  = 2\ (h_2=h_3=0),\hspace{5mm} 2\ (h_2=h_3=1)
\end{equation}
where $h_2$ and $h_3$ are the numbers of Higgs doublets and triplets ($\in
SU(2)_R$).
The electroweak hypercharge is given by
\begin{equation}
Y\ =\ \textstyle T_3^R+\sqrt{{2\over 3}}T_0
\end{equation}
where $T_3^R$ is a generator of $SU(2)_R$ and $T_0$ is the normalized
generator of $U(1)_{B-L}$. Then we obtain the unification condition
\begin{equation}
\left.{5\over2}\cdot{\alpha_y^{-1}-\alpha_w^{-1}\over
\alpha_w^{-1}-\alpha_c^{-1}}\right|_{\mu=M_I} = R
\end{equation}
from which $M_I$ calculated as
\begin{equation}
M_I\ =\ 1.81\times 10^{10}\ {\rm GeV},
\end{equation}
and the unification mass is given as
\begin{equation}
M_X\ =\ 3.73\times 10^{15}\ {\rm GeV}.
\end{equation}
These numbers are comparable to those obtained by Shaban
and Stirling \cite{ss92}, but our method of testing the
unification is simpler.

For $SO(10)\rightarrow SU(5)_{flipped}\times U(1)$, the situation is
not better than the $SU(5)$ model.  This can be easily understood from
Fig.~2(a) ($h_2=1$) where the crossing point of $\alpha_c^{-1}$ and
$\alpha_w^{-1}$ is higher than $\alpha_1^{-1}$.  Let the mass scale of
crossing point of $SU(3)_c$ and $SU(2)_w$ couplings be $M_I$.
At $M_I$, $\alpha_1^{-1}$ jumps slightly due to the mixing
$\alpha_y^{-1}=(24/25)\alpha_1^{-1}+(1/25)\alpha_5^{-1}$, but this
shift is not enough to overcome $\alpha_5^{-1}$.  Therefore,
there is no possibility of further unification of $SU(5)_{flipped}
\times U(1)$ above $M_I$ with $h_2=1$.

As a final example, let us consider $SU(N)$ family unification models.
There are many varieties for hypercharge assignments, but we will focus
on the simplest generalization\cite{family},
\begin{equation}
Y\ =\ {\rm diag} (-{1\over 3},\ -{1\over 3},\ -{1\over 3},\ q,\
\frac{1}{2},\ \frac{1}{2},\ -q,\ \cdots )
\end{equation}
where $\cdots$ are zeros, $SU(3)_c$ is embedded in the $3\times 3$
square matrix
of the first three rows and columns, and $SU(2)_w$ is embedded in the
$2\times 2$ square matrix of
the fifth and sixth rows and columns.
More general hypercharge assignment is
possible with $\Sigma_i q_i=0$ for $q_i$ are the diagonal entries
of $Y$ beyond $SU(5)$.  The simplest example is given for $SU(7)$.
Thus let us focus on $SU(7)$.
The symmetry breaking pattern is assumed to be
\begin{eqnarray}
SU(7) & \stackrel{M_X}{\longrightarrow} &
        SU(4)\times SU(3)\times U(1) \nonumber\\
      & \stackrel{M_I}{\longrightarrow} &
        SU(3)_C\times SU(2)_L\times U(1)_Y
\end{eqnarray}
where $SU(4)$ is embedded in the $4\times 4$ square matrix of the first
four rows and columns and $SU(3)$ is embedded in the $3\times 3$ square
matrix of the next three rows and columns.
Then,
\begin{equation}
R = \frac{11-\frac{1}{14}h_3}{\frac{11}{3}+\frac{1}{6}h_3}
  = 3(h_3=0),\hspace{5mm} 2.71(h_3=2),\hspace{5mm} 2.09(h_3=8)
\end{equation}
where $h_3$ is the number of Higgs triplets (which will become
eventually the number of Higgs doublets of $SU(2)_L$) split from
fundamental representations.
As simple examples of hypercharge assignments,
let us consider $q=0$, $q=1/3$ and $q=1/2$.
We obtain the following electroweak hypercharge generator $Y$ and $r(\mu)$
for each case,
\begin{eqnarray}
q=0\ :\ \ Y &=& \textstyle
-\sqrt{{7\over 6}}T_0+\sqrt{{1\over 3}}T_8-\sqrt{{1\over 6}}T_{15} \nonumber\\
r(\mu) &=& {10\over7}\cdot{\alpha_y^{-1}-\alpha_w^{-1}\over\alpha_w^{-1}
-\alpha_c^{-1}}+{1\over7}\\
q={1\over3}\ :\ \ Y &=& \textstyle
-\sqrt{14\over27}T_0+\sqrt{25\over 27}T_8-\sqrt{2\over 3}T_{15} \nonumber\\
r(\mu) &=& {{45\over14}\alpha_y^{-1}-{39\over 14}\alpha_w^{-1}-
{9\over7}\alpha_c^{-1}\over \alpha_w^{-1}-\alpha_c^{-1}}\\
q={1\over 2}\ :\ \ Y &=& \textstyle
-\sqrt{{7\over 24}}T_0+\sqrt{{4\over 3}}T_8-\sqrt{{25\over24}}T_15 \nonumber\\
r(\mu) &=& {{40\over7}\alpha_y^{-1}-{39\over7}\alpha_w^{-1}
-{25\over7}\alpha_c^{-1}\over\alpha_w^{-1}-\alpha_c^{-1}}
\end{eqnarray}
\begin{table}
\caption{
\it Several extensions of the standard model for unification.  $h$ is
the number of Higgs doublets needed.  For unification, the requirement
of supersymmetry or superstring are also shown.
Cases for $M_X<10^{15}$ GeV are forbidden from the proton decay experiments.
Cases $M_I=M_Z$ correspond to no intermediate mass scale.}
\begin{center}
\begin{tabular}{|c|c|c|c|c|c|} \hline
GUT group&Susy&String&$h$&$M_I\ $[GeV]&$M_X\ $[GeV]
 \\ \hline
$SU(5)$           & No& No&8&            $M_Z$ &$3.4\times10^{13}$ \\ \hline
$SU(5)$           &Yes& No&2&            $M_Z$ &$1.0\times10^{16}$ \\ \hline
$SU(5)\times U(1)$&Yes&Yes&2&            $M_Z$ &$1.0\times10^{16}$ \\ \hline
$SO(10)\rightarrow$ LR model
                  & No& No&2&$1.8\times10^{10}$&$3.7\times10^{15}$ \\ \hline
$SU(7)\ (c=1/3)$  & No& No&2&$9.6\times10^{ 6}$&$4.6\times10^{16}$ \\ \hline
$SU(7)\ (c=1/2)$  & No& No&2&$1.1\times10^{ 6}$&$4.2\times10^{16}$ \\ \hline
\end{tabular}
\end{center}
\end{table}
To see the unification condition explicitly,
we show the $r(\mu)$ and $R$ plot in Fig.~4(a) for a few $q$'s
and for a few $h_3$'s.
For $q=0$, the unification is possible when $h_3\ge8$. But then the proton
lifetime constraint is not satisfied.
For $q=1/3$ and $q=1/2$, the unification is possible for any plausible value
of $h_3$.  In Fig.~4(b), we present the evolution of coupling constants
for $q=1/2$.
The values of $M_I$ and $M_X$ for $h_3=2$ are
\begin{eqnarray}
q={1\over 3} &:& \ M_I=9.6\times 10^6\ {\rm GeV},\ \ M_X=4.6\times 10^{16}
\ {\rm GeV} \\
q={1\over 2} &:& \ M_I=1.1\times 10^6\ {\rm GeV},\ \ M_X=4.2\times
10^{16}\ {\rm GeV}
\end{eqnarray}

In $SU(7)$, two generations can be accommodated in the spinor representation
\cite{family}
of $SO(14)$, e.g. ${\bf1}+{\bf7}^*+{\bf21}+{\bf35}^*$.
To include the third generation,
one must repeat the spinor representation.  Other representations can be
used for realistic unifications in $SU(7)$.

The various possibilities for coupling constant unification studied
in this paper are summarized in Table~1.

In conclusion, we showed the possibilities of coupling constant
unification by extending the minimal standard model, either
by introducing superpartners or by introducing an intermediate
mass scale.

\begin{center}
{\large\bf Acknowledgments}
\end{center}

This work is supported in part by the Center for Theoretical Physics,
Seoul National University and by KOSEF--DFG Collaboration program.

\newpage
\bibliographystyle{unsrt}
\def\prl#1#2#3{Phys.\ Rev.\ Lett.\ #1 (#2) #3}
\def\prd#1#2#3{Phys.\ Rev.\ D#1 (#2) #3}
\def\npb#1#2#3{Nucl.\ Phys.\ B#1 (#2) #3}
\def\plb#1#2#3{Phys.\ Lett.\ B#1 (#2) #3}
\def\opcit#1#2#3{#1 (#2) #3}

\end{document}